\documentclass[sigconf]{acmart}

\usepackage{booktabs} 
\usepackage{balance}
\usepackage{xcolor}
\usepackage{lipsum}
\setcopyright{none}
\setcopyright{acmcopyright} 
\begin{document}


\title{A Tool to Extract Structured Data from GitHub}

\author{Shreyansh Surana}
\affiliation{DAIICT Gandhinagar, India}
\email{201601112@daiict.ac.in }

\author{Smit Detroja}
\affiliation{DAIICT Gandhinagar, India}
\email{201601113@daiict.ac.in}

\author{Saurabh Tiwari}
\affiliation{DAIICT, Gandhinagar, India}
\email{saurabh\_t@daiict.ac.in}

\begin{abstract}

GitHub repositories consist of various detailed information about the project contributors, the number of commits and its contributors, releases, pull requests, programming languages, and issues. However, no systematic dataset of open source projects exists which features detailed information about the repositories on GitHub for knowledge acquisition and mining. In this paper, we developed tool support, named \textit{GitRepository}, which helps in creating a data-set of repositories based on the proposed schema. Out of initial 1680 repositories, the dataset hosts 620 repositories (with applied basic filters of stars and forks), and 247 repositories (after applying all pre-defined filters). The tool extracts the information of GitHub repositories and saves the data in the form of CSV. files and a database (.DB) file. 

\end{abstract}


\maketitle

\section{Introduction}
Open Source Development (OSD) is now increasing with a rapid release rate, consequently, increases the number of GitHub repositories. GitHub is the largest OSD platform which provides version control using Git and runs in the command line interface. It integrates many features of social network coding such as forks, giving stars to the repository, making the repository private and so on. Therefore, the choice for developers from large companies too~\cite{casalnuovo2015developer}\cite{liu2018recommending}. GitHub can be easily connected to many project management tools such as Amazon and Google Cloud accounts, and Android Studio. The main reason for it being so much widespread is that it allows multiple collaborators (e.g., developers, testers, analysts etc.) to work on a single project at the same time. A single command can make sure that all the collaborators are on the same page.

Due to the large use of the Git platform, a large amount of data has been generated with multiple and rapid releases. The motivation behind this work is to understand and filter out a huge amount of data using some predefined filters. The data can help find trending developers, languages, and repositories across the platform. The data can be useful to identify developers and also finding patterns in selecting the language for a project, issues related to that, the relation between contributors and watchers. To collect such data, we developed a tool which helps in fetching data from GitHub. Subsequently, the data can be analysed to identify (or know) the relation between watchers and contributors, the relation between stars and forks, patterns of contributions among the GitHub users, the relation between programming languages and issues of the repository.

In this paper, we have developed tool support to extract repositories along with the information related to contributors, issues, pull requests, releases, and subscribers. The tool provides three different options: creating a data set of repositories, creating consolidated data files of individual users and individual repository. The basic terminologies used in the paper are, a repository (a storage space where the project is live), forks (copying the repository to our own space for personal use), and stars (for following a specific repository to know about future developments and changes). The tool uses GitHub API\footnote{https://developer.github.com/v3/} in the back end, with the help of python library requests, to get all the relevant data needed for creating the data-set. The tool uses a Tkinter python library\footnote{https://docs.python.org/3/library/tkinter.html} to create a UI so that the tool can be used without any hassle.

\section{Related Work}
Numerous studies have been conducted to systematically generate the data from GitHub and subsequent analysis. However, dealing with a huge large set of repositories is difficult. This is because several Git repositories are created and later not used for any purpose. Several frameworks (e.g., dataset mirroring GitHub’s data (i.e., GHTorrent~\cite{ghtorrent}\cite{Gousi13}, GitHub Archive\footnote{https://www.gharchive.org/}, BOA~\cite{BOA}), GitHub API\footnote{https://developer.github.com/}, GitHub Search API~\footnote{https://developer.github.com/v3/search/} and a mixture of the previous ones~\cite{cseker2020summarising} are also available in the literature which provides offline mirroring of data through the GitHub API. These data sets offer data in different formats in which one can execute queries and fetch the data. For instance, GHTorrent offers a queriable offline mirror of data that stores the raw JSON responses to a MongoDB database~\cite{Gousi13}.

GHTorrent~\cite{ghtorrent} gets all the public content from the GitHub in the real-time and adds to its already huge database. It is also storing the data in the JSON format other than database formats. Their data has everything related to the GitHub due to it being a real-time service. The dataset that our tool will collect would always be part of the subset of the GHTorrent. But to get to that subset would require to write database queries and execute them on a large database (more than 6.5 billion rows) or execute them on multiple instances of databases, which can be a very costly operation. GHArchive monitors the GitHub public event timeline, archives those events, and recursively crawl and archive their contents and dependencies. Those archives will then be made available for download on a daily or monthly basis. 

GHArchive is same as GHTorrent, and the difference is in terms of storing of the data (GHTorrent is doing it so that data will be available from past even after it has been deleted and GHArchive is doing it so that it can store the data for future use (even after 1000 years)). BOA~\cite{BOA} is an infrastructure to fetch data on huge projects efficiently. They used hardware and other techniques to speed up the process and make it more scalable. It reduces the programming efforts as the user don't have to explicitly parallelise their code.

Rapid Release~\cite{b4} is also tool support developed for creating a dataset of GitHub repositories. The proposed tool has several similarities with the Rapid Release and can be considered an extension of the Rapid Release. The similarities and differences with the Rapid Release tool are as follows:

\noindent \textit{Similarities}
\begin{itemize}
    \item Creating dataset of GitHub repositories 
    \item The proposed schema is inspired by the Rapid Release (we also fetched pull requests data and subscribers data which is the future scope of Rapid Release)
    \item Data is available in the database (.db) and, comma-separated values (.CSV) format, which can be directly used by anyone for analysis or viewing the data.
\end{itemize}

\noindent \textit{Differences}
\begin{itemize}
    \item Rapid Release dataset consists of repositories which have new releases every 5-35 days whereas the repositories fetched in our tools are based on inputs given by the user.
    \item Dataset is fixed in Rapid Release, whereas \textit{GitRepository} creates a new data-set according to the pre-defined several parameters.
    \item \textit{GitRepository} is capable of fetching data related to a single user and a single repository. The private repositories can also be fetched.
    \item The data on Rapid Release was fixed on releases whereas the \textit{GitRepository} dataset includes generic terms of GitHub.
    \item With some knowledge of Python, the user can tweak the dataset, removing or adding information according to their needs. Data other than the provided schema is hierarchically stored in arrays.
\end{itemize}

\section{Tool Support for Dataset Creation}
\begin{figure}[htbp]
\centering{\includegraphics[scale=0.79]{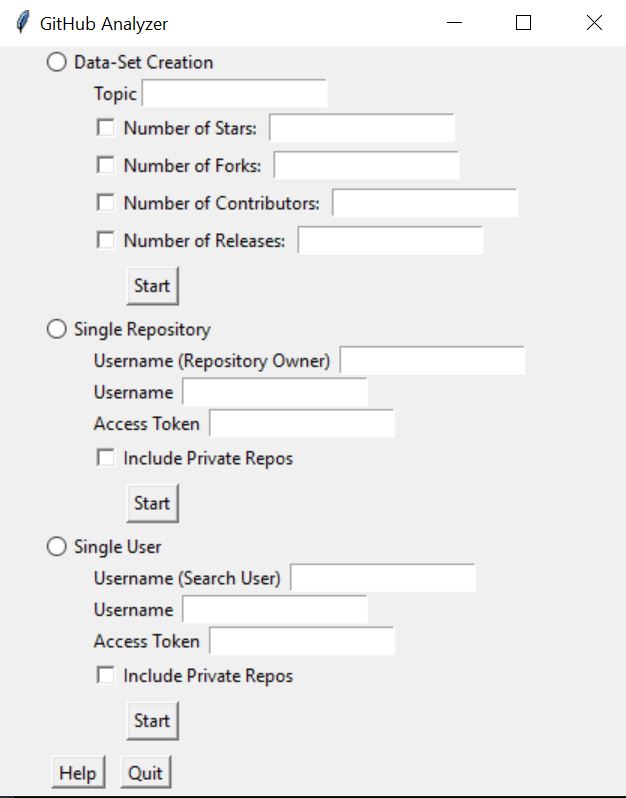}}
\vspace{-3.4ex}
\caption{Snapshot of the \textit{GitRepository} Tool}
\vspace{-1.6ex}
\label{figToolSnap}
\end{figure}

The \textit{GitRepository} helps in creating a dataset of multiple repositories, single repository, and single-user statistics. Specifically, multiple repositories dataset consists of repositories belonging to multiple users. This means that it takes up the data of repositories from many users that satisfy the search criteria, and creates a dataset. Whereas, a single repository means that it has only one owner, and the data would be only about that specific repository instead of multiple repositories. Single user data-set implies that the data-set would have all the information, including all the repositories, regarding a single user. Figure~\ref{figToolSnap} shows the layout of the tool which is primarily divided into three components:
\begin{itemize}
    \item Searching GitHub for the repositories related to a specific domain and get all the possible details such as details of contributors, commits, releases, pull requests, languages and issues. To generate the dataset, the name of the topic (domain), and filter details such as the minimum number of stars, forks, releases and contributors need to be specified in the tool.
    \item Getting all the details of a single repository with its commit details. For this, the tool takes two inputs: username of the owner of a repository and an access token of the same username\footnote{https://help.github.com/en/github/authenticating-to-github/creating-a-personal-access-token-for-the-command-line}.
    \item Getting all the details about a GitHub user. For this, the user needs to specify the username of the owner of a repository and an access token of the same username$^3$.
\end{itemize}

\subsection{Features, Source Code \& Demonstration}
The salient features of the \textit{GitRepository} are as follows:
\begin{itemize}
    \item Creating a data-set of multiple repositories, in CSV files and DB file, based on a specific domain and filters provided by the user of the tool.
    \item Showing percentage wise contribution of all the contributors of a single repository.
    \item Storing all the user information, including details about all repositories in CSV files.
    \item Showing the most prominent programming languages in the user's profile using percentage wise programming languages used in the user's all repositories.
\end{itemize}

The video demonstration of the \textit{GitRepository} can be found at \textit{\url{https://www.youtube.com/watch?v=9ptff60m1ig}}. The source code of the tool along with other details is available on Git - \textit{\url{https://github.com/shreyansh08/GithubAnalyzer}}.

\subsection{Creation of the Dataset}
\begin{figure}[!ht]
\centering{\includegraphics[height=3.8in,width=3.4in,angle=0]{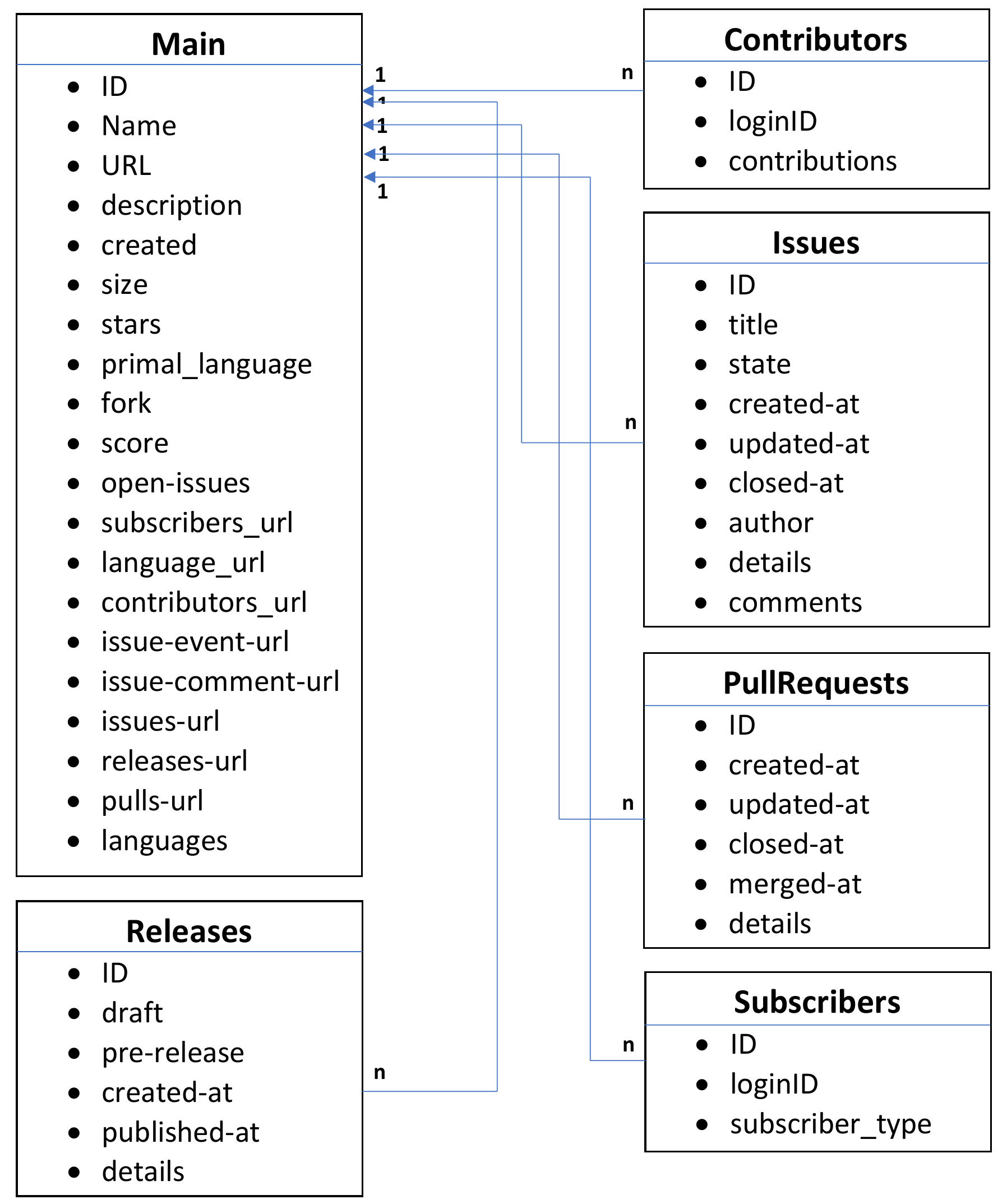}}
\vspace{-2.5ex}
\caption{Proposed schema for dataset creation - \textit{GitRepository}}
\label{figSchema}
\end{figure}

A user needs to set multiple parameters to create a data set. User needs to give a topic name and can set the minimum number of stars, contributors, releases and forks that need to be present in the repository to be included in the data set. The dataset is created by following the proposed pre-defined schema is shown in Figure~\ref{figSchema}. 

\noindent The step-wise process for the creation of the dataset is as follows:

\textit{Step I}: We searched the repositories that a user wants using GitHub API and sorted it with respect to the number of stars of the repository. The search criteria are determined by the user where he/she decided the topic name (domain) to be searched and decide on threshold number of stars the repository must have and the minimum number of times the repository must have been forked.

\textit{Step 2}: In this step, we take each of the repositories and determine the prominent languages of each repository. This step enables us to collect the data regarding the languages that are used in a specific domain/field that has been searched. All the languages which are not prominent are also collected and saved. We also apply the second filter were, if user has given minimum number of releases, it is used here to filter out the repositories which have releases less than threshold minimum value given by the tool user.

\textit{Step 3}: In this step, we extract more details about the repository, including details regarding the pull requests and collaborators of the project and filter the repositories based on inputs given by the user during the initial stage. We extract the issues that each repository has encountered. Using GitHub API, we collect all the details of each issue including the comments and details of the issue. This helps us in analysis questions that we have discussed further. We also fetch all the pull requests generated for the repository, and store it.

\begin{figure}[htbp]
\centering{\includegraphics[height=3.67in,width=3.4in,angle=0]{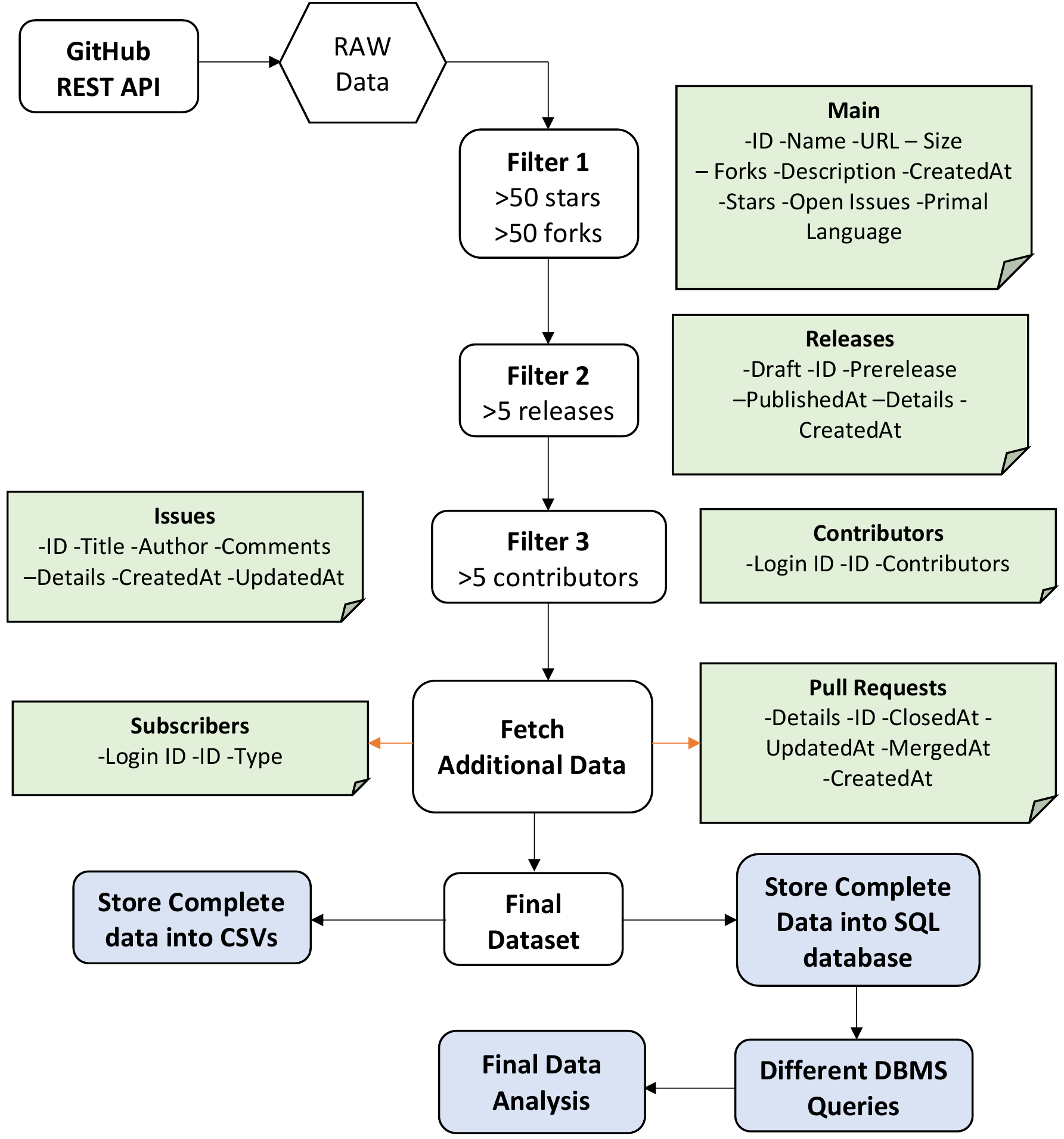}}
\vspace{-2.5ex}
\caption{\textit{GitRepository} dataset construction and analysis process overview}
\label{figDataFlowofTool}
\vspace{-1.6ex}
\end{figure}

To understand better, the flow of data and the filters applied at a stage can be visualised using Figure~\ref{figDataFlowofTool}.

\begin{figure*}[!ht]
\centering{\includegraphics[scale=0.7]{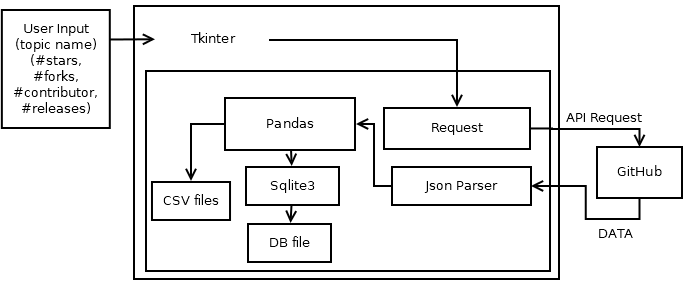}}
\vspace{-1.6ex}
\caption{Architecture of \textit{GitRepository}}
\vspace{-1.6ex}
\label{figflow}
\end{figure*}

\subsection{Single User and Single Repository}
Similar to above process, using GitHub API, we gave user of the tool more options to get all the data about a specific single repository and to get all the data about a specific GitHub user. A single repository analysis fetches all the basic details about the repository with all the commits whereas a single user profile search fetches all the data about the user. 

\section{Architecture Of GitRepository}
\textit{GitRepository} is an interactive tool developed in Python 3.7\footnote{https://www.python.org/}. The tool uses Requests and HTTPBasicAuth library to fire the API queries to GitHub using the REST API\footnote{https://www.restapitutorial.com/} for data extraction and uses JSON\footnote{https://www.json.org/} file parser and Pandas\footnote{https://pandas.pydata.org/} Data Frame to organize and store the data. Pandas' feature of converting data frames to csv is used to create CSV files. Python library SQLite3\footnote{https://www.sqlite.org/index.html} is used to create a database out of the given formatted data (The database file can be properly opened only in SQLite3 software). Jupyter\footnote{https://jupyter.org/} Notebook is used for detailed analysis of the data collected by the tool. Library MatPlotLib\footnote{https://matplotlib.org/}, WordCLoud\footnote{https://www.wordclouds.com/} and Seaborn\footnote{https://seaborn.pydata.org/} is used to showcase the important results of the tool. Git\footnote{https://git-scm.com/}, GitPython\footnote{https://pypi.org/project/GitPython/} and Pydriller\footnote{https://pydriller.readthedocs.io/en/latest/} is used to analyze the local repositories on the computer. The UI for ease of use is created using the help of python library Tkinter\footnote{https://docs.python.org/3/library/tkinter.html}. Tkinter allows to create an interactive UI, and minimizes the use of directly running the main source file. Figure~\ref{figflow} shows the architecture of the tool \textit{GitRepository}.

\section{Summary \& Future Prospects}
In this paper, we have developed tool support, \textit{GitRepository}, for creating dataset about the repositories from the GitHub. The tool follows a proposed schema for creating the dataset. The dataset takes parameters such as the number of stars, fork, contributors and releases as input to fetch the data. The tool also able to fetch data for a single repository and a single user. 

The proposed tool doesn't allow users to add a new parameter for the creation of dataset, hence it can be extended to incorporate user's preferences before creating the dataset. Similar to the queries for extracting data, we generate percentage-wise contribution of a contributor to a repository, more queries can be written to get ideas about different parts of the repository like which user reports the most number of issues or how frequent the pull requests are merged or what is the time difference between different releases of the product.

Since the dataset created is limited to the main topic, it would be interesting to identify and generate various insights about the domain. For example, getting the programming languages used in a specific domain; getting the most common issues present in various domains. If the dataset consists of repositories belonging to a specific language, then it would be interesting to identify the prominent issues recurring in all those repositories and relating it with the programming language. This can help the language creators get an idea about what are the common challenges that their programming language face.


\begin{thebibliography}{7}


\ifx \showCODEN    \undefined \def \showCODEN     #1{\unskip}     \fi
\ifx \showDOI      \undefined \def \showDOI       #1{#1}\fi
\ifx \showISBNx    \undefined \def \showISBNx     #1{\unskip}     \fi
\ifx \showISBNxiii \undefined \def \showISBNxiii  #1{\unskip}     \fi
\ifx \showISSN     \undefined \def \showISSN      #1{\unskip}     \fi
\ifx \showLCCN     \undefined \def \showLCCN      #1{\unskip}     \fi
\ifx \shownote     \undefined \def \shownote      #1{#1}          \fi
\ifx \showarticletitle \undefined \def \showarticletitle #1{#1}   \fi
\ifx \showURL      \undefined \def \showURL       {\relax}        \fi
\providecommand\bibfield[2]{#2}
\providecommand\bibinfo[2]{#2}
\providecommand\natexlab[1]{#1}
\providecommand\showeprint[2][]{arXiv:#2}

\bibitem[\protect\citeauthoryear{Casalnuovo, Vasilescu, Devanbu, and
  Filkov}{Casalnuovo et~al\mbox{.}}{2015}]%
        {casalnuovo2015developer}
\bibfield{author}{\bibinfo{person}{Casey Casalnuovo}, \bibinfo{person}{Bogdan
  Vasilescu}, \bibinfo{person}{Premkumar Devanbu}, {and}
  \bibinfo{person}{Vladimir Filkov}.} \bibinfo{year}{2015}\natexlab{}.
\newblock \showarticletitle{Developer onboarding in GitHub: the role of prior
  social links and language experience}. In \bibinfo{booktitle}{{\em
  Proceedings of the 2015 10th joint meeting on foundations of software
  engineering}}. \bibinfo{pages}{817--828}.
\newblock


\bibitem[\protect\citeauthoryear{Dyer, Nguyen, Rajan, and Nguyen}{Dyer
  et~al\mbox{.}}{2013}]%
        {BOA}
\bibfield{author}{\bibinfo{person}{Robert Dyer}, \bibinfo{person}{Hoan~Anh
  Nguyen}, \bibinfo{person}{Hridesh Rajan}, {and} \bibinfo{person}{Tien~N.
  Nguyen}.} \bibinfo{year}{2013}\natexlab{}.
\newblock \showarticletitle{Boa: A Language and Infrastructure for Analyzing
  Ultra-Large-Scale Software Repositories}. In \bibinfo{booktitle}{{\em
  Proceedings of the 2013 International Conference on Software Engineering}}
  {\em (\bibinfo{series}{ICSE ’13})}. \bibinfo{publisher}{IEEE Press},
  \bibinfo{pages}{422–431}.
\newblock
\showISBNx{9781467330763}


\bibitem[\protect\citeauthoryear{Gousios}{Gousios}{2013}]%
        {Gousi13}
\bibfield{author}{\bibinfo{person}{Georgios Gousios}.}
  \bibinfo{year}{2013}\natexlab{}.
\newblock \showarticletitle{The GHTorrent dataset and tool suite}. In
  \bibinfo{booktitle}{{\em Proceedings of the 10th Working Conference on Mining
  Software Repositories}} {\em (\bibinfo{series}{MSR '13})}.
  \bibinfo{publisher}{IEEE Press}, \bibinfo{address}{Piscataway, NJ, USA},
  \bibinfo{pages}{233--236}.
\newblock
\showISBNx{978-1-4673-2936-1}
\showURL{%
\url{http://dl.acm.org/citation.cfm?id=2487085.2487132}}


\bibitem[\protect\citeauthoryear{Gousios and Spinellis}{Gousios and
  Spinellis}{2012}]%
        {ghtorrent}
\bibfield{author}{\bibinfo{person}{Georgios Gousios} {and}
  \bibinfo{person}{Diomidis Spinellis}.} \bibinfo{year}{2012}\natexlab{}.
\newblock \showarticletitle{GHTorrent: GitHub’s Data from a Firehose}. In
  \bibinfo{booktitle}{{\em Proceedings of the 9th IEEE Working Conference on
  Mining Software Repositories}} {\em (\bibinfo{series}{MSR ’12})}.
  \bibinfo{publisher}{IEEE Press}, \bibinfo{pages}{12–21}.
\newblock
\showISBNx{9781467317610}


\bibitem[\protect\citeauthoryear{Joshi and Chimalakonda}{Joshi and
  Chimalakonda}{2019}]%
        {b4}
\bibfield{author}{\bibinfo{person}{Saket~Dattatray Joshi} {and}
  \bibinfo{person}{Sridhar Chimalakonda}.} \bibinfo{year}{2019}\natexlab{}.
\newblock \showarticletitle{RapidRelease-A Dataset of Projects and Issues on
  Github with Rapid Releases}. In \bibinfo{booktitle}{{\em 2019 IEEE/ACM 16th
  International Conference on Mining Software Repositories (MSR)}}. IEEE,
  \bibinfo{pages}{587--591}.
\newblock


\bibitem[\protect\citeauthoryear{Liu, Yang, Zhang, Ray, and Rahman}{Liu
  et~al\mbox{.}}{2018}]%
        {liu2018recommending}
\bibfield{author}{\bibinfo{person}{Chao Liu}, \bibinfo{person}{Dan Yang},
  \bibinfo{person}{Xiaohong Zhang}, \bibinfo{person}{Baishakhi Ray}, {and}
  \bibinfo{person}{Md~Masudur Rahman}.} \bibinfo{year}{2018}\natexlab{}.
\newblock \showarticletitle{Recommending github projects for developer
  onboarding}.
\newblock \bibinfo{journal}{{\em IEEE Access\/}}  \bibinfo{volume}{6}
  (\bibinfo{year}{2018}), \bibinfo{pages}{52082--52094}.
\newblock


\bibitem[\protect\citeauthoryear{{\c{S}}eker, Diri, and Arslan}{{\c{S}}eker
  et~al\mbox{.}}{2020}]%
        {cseker2020summarising}
\bibfield{author}{\bibinfo{person}{Abdulkadir {\c{S}}eker},
  \bibinfo{person}{Banu Diri}, {and} \bibinfo{person}{Halil Arslan}.}
  \bibinfo{year}{2020}\natexlab{}.
\newblock \showarticletitle{Summarising Big Data: Common GitHub Dataset for
  Software Engineering Challenges}.
\newblock \bibinfo{journal}{{\em arXiv preprint arXiv:2006.04967\/}}
  (\bibinfo{year}{2020}).
\newblock


\end{thebibliography}


\end{document}